\documentclass[prl,aps,secnumarabic,showpacs,twocolumn]{revtex4}
\usepackage{bm,latexsym,amsmath,amsfonts,graphicx,color}
\usepackage{hyperref,url,hhline,multirow}
\usepackage[usenames,dvipsnames,svgnames,table]{xcolor}

\usepackage{url}
\usepackage{hyperref}

\definecolor{gesfpurple}{rgb}{0.47,0.19,0.42}

\definecolor{gesflanse}{rgb}{0.00,0.50,0.50}

\definecolor{gesfblue}{rgb}{0.08,0.42,0.76}

\definecolor{gesfred}{rgb}{1,0,0}

\definecolor{gesfwhite}{rgb}{1,1,1}

\definecolor{gesfblack}{rgb}{0,0,0}

\newcommand{\geqn}[1]{\hypersetup{linkcolor=blue}(\ref{#1})\hypersetup{linkcolor=blue}}
\newcommand{\gfig}[1]{{\hypersetup{linkcolor=violet}Fig.~\ref{#1}\hypersetup{linkcolor=blue}}}

\allowdisplaybreaks[4]

\begin{document}


\title{\Large Atmospheric Trident Production for Probing New Physics}
\author{
Shao-Feng Ge\footnote{gesf02@gmail.com}, 
Manfred Lindner\footnote{lindner@mpi-hd.mpg.de},
Werner Rodejohann \footnote{werner.rodejohann@mpi-hd.mpg.de}
}
\affiliation{
Max-Planck-Institut f\"{u}r Kernphysik, Heidelberg 69117, Germany
}

\date{\today}

\begin{abstract}
\noindent
We propose to use atmospheric neutrinos as a powerful probe of new physics
beyond the Standard Model via neutrino trident production. 
The final state with double muon tracks
simultaneously produced from the same vertex is a distinctive signal at
large Cherenkov detectors.
We calculate the expected event numbers of trident
production in the Standard Model. 
To illustrate the potential of this process to probe new physics we 
obtain the sensitivity 
on new vector/scalar bosons with coupling to muon and tau neutrinos.
\end{abstract}

\maketitle

{\it \bf Introduction} -- 
The neutrino oscillation is the first place that new physics (NP)
beyond the Standard Model (SM) was observed \cite{sk}. 
In less than two decades, a consistent minimal picture has emerged: three flavor
states whose mixing with each other is described by
a unitary matrix, together with interactions predicted by the SM \cite{nuRev}.
However, neutrino oscillation cannot directly test the underlying
picture behind the neutrino mass and mixing.
We need new concept to go beyond.

\begin{figure}[b]
\centering
\includegraphics[width=.45\textwidth]{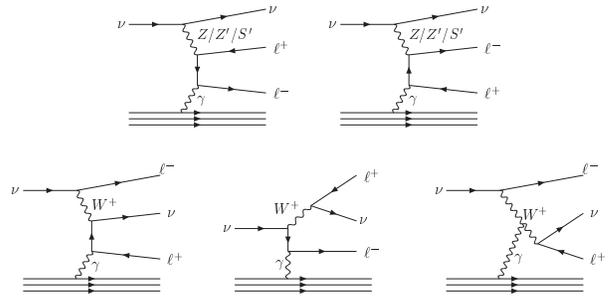}
\caption{Feynman diagrams for neutrino trident production in the SM and in the presence of 
a vector (scalar) boson $Z'$ ($S'$). }
\label{fig:feyn}
\end{figure}

The rare process of neutrino trident production \cite{trident,coherent},
\begin{equation}
  \stackrel{(-)}{\nu_\ell} N
\to
  \stackrel{(-)}{\nu_\ell} \ell^+ \ell^- X \mbox{ with } \ell = e,\mu,\tau\,,
\label{eq:nuTrident}
\end{equation}
can serve the purpose of directly probing the underlying principle of neutrino physics.
While neutrino oscillation experiment focuses on reconstructing the initial state,
namely the incident neutrino energy and flavor, neutrino trident production can
produce new particles as intermediate state (see \gfig{fig:feyn}) and
constrain their properties by observing the final-state particles.
Essentially, we can turn an oscillation experiment into a
``{\it neutrino collider}''.

{\it \bf Neutrino Trident Production in the Standard Model} --
In neutrino trident production, three leptonic particles appear in the final state. One is a 
neutrino and the other two a pair of leptons with opposite charge,
as plotted in \gfig{fig:feyn}. 

At leading order, the leptonic part connects with 
nuclei through a photon. 
The $Z$ boson can also establish the connection but its contribution 
is highly suppressed by its heavy propagators. 
Suppression also occurs in the last diagram in \gfig{fig:feyn} through two $W$ propagators.
Since the connection with nuclei is established by a photon, which is electrically neutral,
there is no difference between the neutrino and anti-neutrino modes, i.e.\ 
$\sigma_\nu(E) = \sigma_{\bar \nu}(E)$. However, the differential distributions 
of the two final-state charged leptons have some measurable difference if 
charge identification is possible. 
Because of the $W$ contribution, there is no interchange symmetry between the two charged 
leptons.
Since the connection with the nucleus is established through a photon, this process actually 
probes the electromagnetic structure \cite{Gao:2003ag} of the target nuclei
with suppressed momentum transfer. 
We implement the form factors for coherent \cite{coherent} and diffractive
\cite{diffractive} regions in CompHEP \cite{CompHEP} to simulate the
$\nu N \rightarrow \nu \ell^+ \ell^- N$ process with four-body final state.
\begin{figure}[h]
\centering
\includegraphics[height=4.2cm,width=3.62cm,angle=-90]{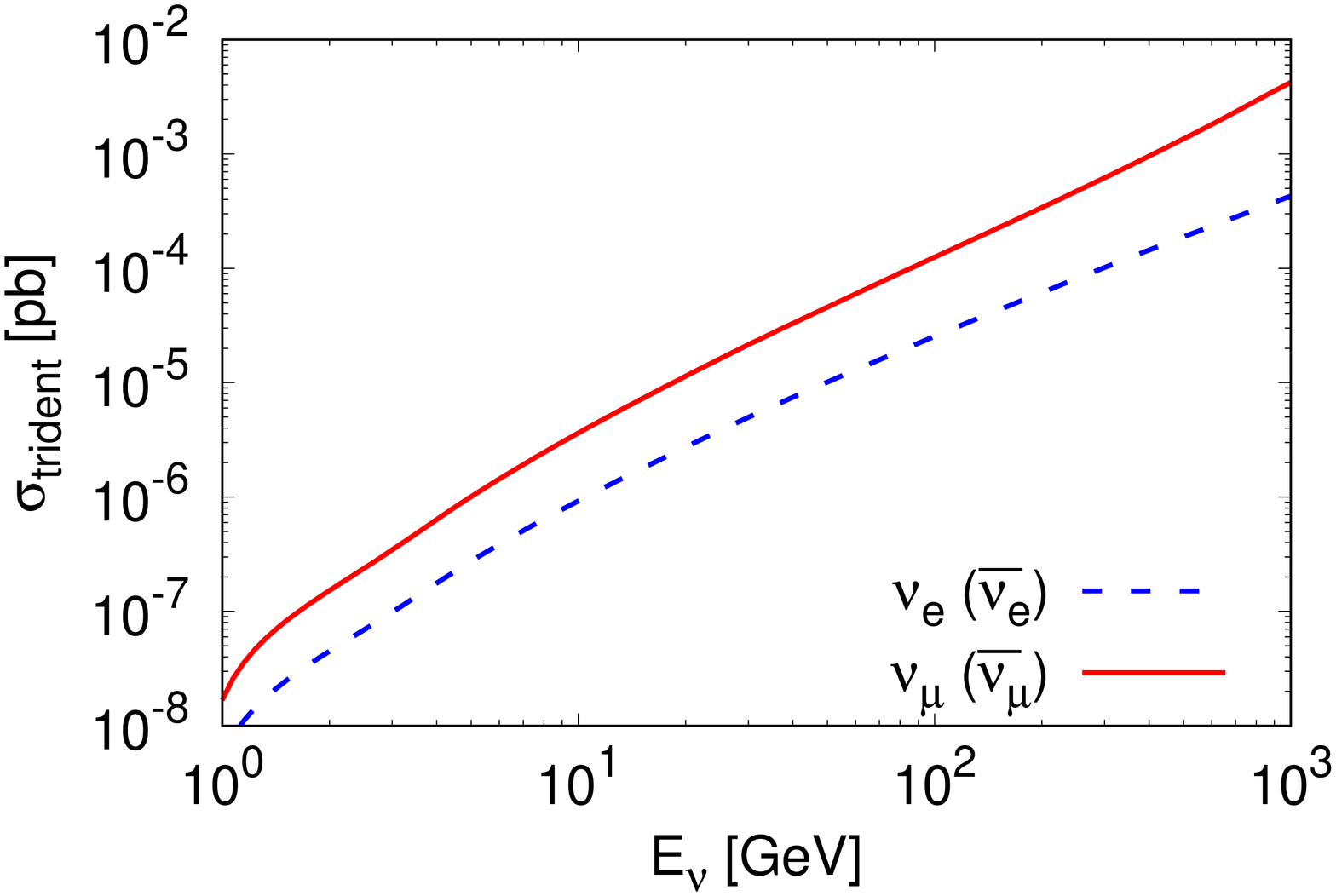}
\includegraphics[height=4.2cm,width=3.62cm,angle=-90]{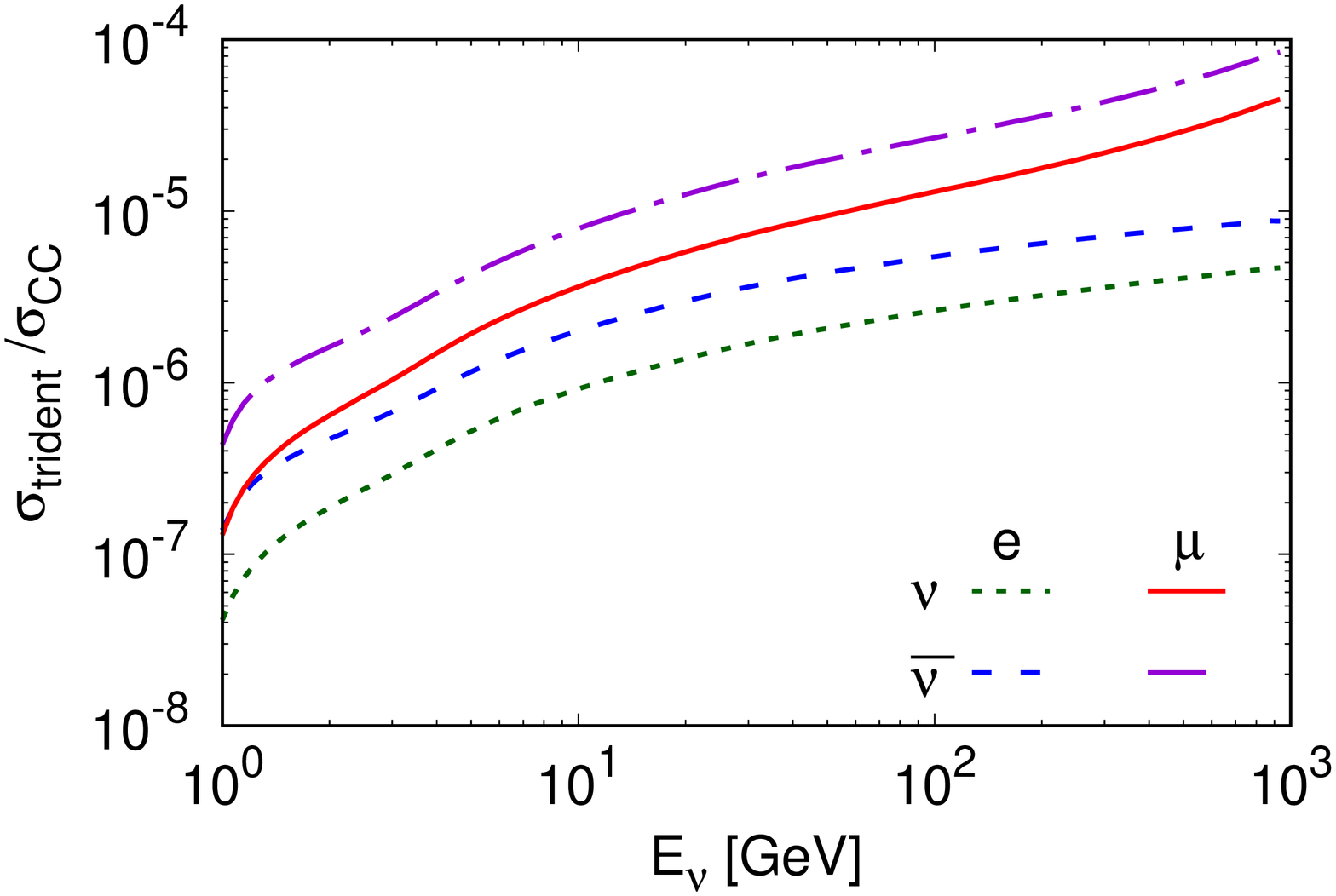}
\caption{The SM prediction of the neutrino trident production cross section on water target (left) and its relative
         value with respect to the cross section of CC scattering (right).}
\label{fig:xsec}
\end{figure}

The energy dependence of the trident production cross section $\sigma_{\rm trident}$ is shown in
\gfig{fig:xsec}. Below 2\,GeV, $\sigma_{\rm trident}$ increases
quickly after the channel opens up at the muon pair production threshold
$E^{\rm th}_\nu = 2 m_\mu \simeq 210 \, \mbox{MeV}$. For comparison, the cross section 
$\sigma_{\rm CC}$ of the charged-current (CC) scattering of neutrinos on the same water target is simulated with
GENIE \cite{GENIE}. As shown in the right panel of \gfig{fig:xsec}, the cross section
of neutrino trident production is roughly $4\sim6$ orders smaller than the one of
neutrino scattering in the range $E_\nu \gtrsim 2 \, \mbox{GeV}$. The higher the neutrino
energy, the larger $\sigma_{\rm trident}$ and $\sigma_{\rm trident} / \sigma_{\rm CC}$. 

{\it \bf Neutrino Trident Observation at Cherenkov Detectors} --
To collect a handful of neutrino trident events, the same experiment should be
able to collect at least $\mathcal O(10^6)$ of the ordinary neutrino
CC events which is a quite large event number for the existing and next-generation
neutrino experiments.
Regarding man-made neutrino sources, the only possibility is provided by accelerator experiments \cite{Altmannshofer:2014pba}, 
including the past experiments CHARM-II \cite{CHARM2}, CCFR \cite{CCFR} and NuTeV \cite{NuTeV}, 
as well as the future high-intensity neutrino facilities such has DUNE \cite{Magill:2016hgc}. Here we propose using
the free source of atmospheric neutrinos to observe neutrino trident production at large
Cherenkov detectors, such as PINGU \cite{PINGU} and ORCA \cite{KM3net}.

These Cherenkov detectors are huge ice/water cubes filled with vertical strings of digital optical
modules (DOMs). The ice/water is used both as scattering target
and measurement medium. Charged particles traveling faster than the speed of light in 
ice/water, which is roughly $3/4$ of the speed of light in vacuum, produce Cherenkov radiation and
the DOMs record the energy and arriving time of Cherenkov photons to
reconstruct the momentum and flavor of the charged particles. Especially muons can leave a clear track inside 
the detector while electrons leave more spherical radiation. This difference can be used to
identify muons effectively.

\begin{figure}[t]
\centering
\includegraphics[height=0.48\textwidth,width=4.5cm,angle=-90]{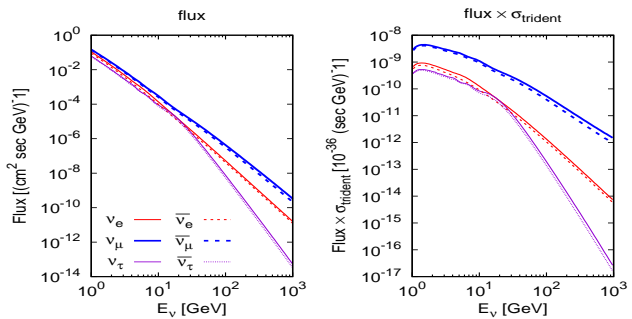}
\caption{Atmospheric neutrino fluxes at South Pole \cite{Honda} after oscillation.
         The Mediterranean sites have similar fluxes.}
\label{fig:flux}
\end{figure}

The original atmospheric neutrino flux \cite{Honda} needs to be folded with neutrino oscillation probabilities, 
 which depend on both oscillation parameters and the Earth matter density. 
A numerical evaluation of neutrino oscillations with decomposition in the propagation basis and 
applying the PREM Earth matter profile \cite{PREM} can be found in \cite{atmos} and has
been implemented in the NuPro package \cite{nupro}. For simplicity, we take the
current global best fit-values \cite{global} of neutrino oscillation parameters to
calculate the fluxes arriving at Cherenkov detectors, as shown in \gfig{fig:flux}.

For a neutrino telescope that cannot distinguish muon charge, such as PINGU
and ORCA, a huge detector size is necessary for the mass hierarchy measurement.
Its effective volume can be as large as $5$\,Mt (PINGU \cite{PINGU}) and $6$\,Mt
(ORCA \cite{KM3net}) at $E_\nu \sim 30 \mbox{\,GeV}$. 
Every year, PINGU can collect $1.6 \times 10^5$ events \cite{Cowen}.
For a period of 6 years, one million events can be collected. This is roughly
what we need to collect around 10 events of neutrino trident production.
The same observation can also happen at the larger versions,
IceCube ($\sim 200\,\mbox{Mt}$) \cite{IceCube},
DeepCore ($\sim 60\,\mbox{Mt}$) \cite{DeepCore},
and ARCA ($\sim 300\,\mbox{Mt}$) \cite{KM3net} at $E_\nu = 500\,\mbox{GeV}$.

\begin{figure}[b]
\centering
\includegraphics[height=0.48\textwidth,width=4.5cm,angle=-90]{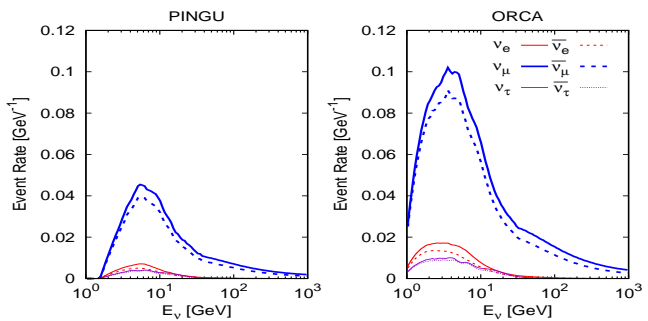}
\caption{The SM prediction of the neutrino trident production event rate at PINGU (left) and ORCA (right).}
\label{fig:rate}
\end{figure}

By folding the cross section of neutrino trident production with the atmospheric neutrino
flux after propagating through the Earth, and taking into account the effective volume,
we calculate the SM prediction for the neutrino trident event rate and display it 
in \gfig{fig:rate} for both neutrinos and anti-neutrinos.
Different from the ordinary scattering process, the event rate of neutrino trident production keeps growing until $E_\nu \sim 10 \, \mbox{GeV}$.

The total rate is 7.4 (17, 63) and 16 (23) events for 10 years running of PINGU (DeepCore,
IceCube) and ORCA (ARCA). The South Pole (Mediterranean) combination can collect
87 (39) events and statistically the uncertainty
can be as small as $11\%$ ($16\%$). Of course, there are systematical uncertainties, such as the
normalization of the predicted atmospheric neutrino flux which can be as large as 20\% \cite{Honda}.
Fortunately, the ordinary neutrino scattering process can collect millions of events to constrain the
normalization with unprecedented sub-percent precision. 

Coincident muon tracks can fake the muon pair from neutrino trident production. However, 
the cosmic muons can be vetoed by the detectors outside or on the 
surface for PINGU, or in case of ORCA with dedicated kinematic cuts. 
The irreducible background comes from the yearly $N_\mu \equiv 8.6 \times 10^4$ 
CC muon events (at PINGU). By applying coincidence cuts, namely requiring the two muons to appear 
simultaneously within a narrow time window $\Delta t$, the fake rate can be suppressed to 
$C_{N_\mu}^2 \times (\Delta t/T)^2$, where $T = 3.15 \times 10^7$ is the length of a year and 
$C_{N_\mu}^2=N_\mu (N_\mu-1)/2$ is the number of possible combinations with two out of the $N_\mu$ events 
to form a pair of muons as fake signal. 
Less than one background double-track requires $\Delta t \lesssim 500$ s, which is definitely
achievable. In addition, the requirement that both muons come from the same vertex should further
reduce the background rate. With increasing event number, the time window decreases as
$\Delta t \propto 1/N_\mu$. So the same technique can also apply to DeepCore, IceCube,
and ARCA. 

Another possible background comes from the CC scattering with one primary muon
and another faked muon from pion or charm decay. However, this background has
totally different kinematics than the signal. While the background event is
associated with hadronic shower to produce pion or charm, the signal has highly
suppressed momentum transfer to the target nuclei due to the massless photon propagator
and the electromagnetic form factor. In addition, the muon from pion or charm
decay tends to have tiny energy. The signal has purely two energetic muons
which is a very clear signal in Cherenkov detector.

All these points are beneficial for the observation of neutrino trident production
at large Cherenkov detectors. First, the large detector size ensures enough event rate.
Second, the double-track signal is easy to identify \cite{double-track}
without modification of the detector and the search can be carried out simultaneously
with the neutrino oscillation measurement. 
Finally, both systematics and background rate are small.

{\it \bf Neutrino Trident Production with New Physics} -- 
The effect of NP considered here appears by replacing the $Z$ boson with a new
vector ($Z'$) or scalar ($S'$) boson. In principle all flavors contribute to 
Eq.\ \geqn{eq:nuTrident}. Here we focus on muon final states, because of strong
constraints on new physics in the electron sector and because muons are easier
to identify than other flavors. 

For illustration of the vector boson case, we consider the $L_\mu - L_\tau$ model \cite{Lmtmod}
which is anomaly free \cite{Lmt}. The relevant Lagrangian is 
\begin{equation}
  \mathcal L_{Z'}
\equiv
  g_{Z'} Q_{\alpha \beta}
\left[
  \overline L_\alpha \gamma^\mu L_\beta
+ \overline \ell_{R \alpha} \gamma^\mu \ell_{R \beta}
\right] Z'_\mu + h.c.
\end{equation}
where $L_\alpha$ denotes the left-handed lepton doublet and $\ell_R$ the right-handed lepton.  
The $Z'$ couples to the muon and tau flavors with opposite charge, 
$Q_{\alpha \beta} = \mbox{diag} (0, 1, -1)$, where
$(\alpha, \beta) \equiv (e, \mu, \tau)$ denotes the three lepton flavors. Apart from the quantum number
assignment, there are two new parameters, the $Z'$ mass $M_{Z'}$ and coupling constant
$g_{Z'}$. Trident production with gauged $L_\mu - L_\tau$ has been discussed for other types of experiments in \cite{Altmannshofer:2014pba,Kaneta:2016uyt}. We do not consider models in which there is additional coupling of the $Z'$ to quarks \cite{Crivellin:2015mga}. 
Although there are various constraints \cite{Altmannshofer:2014pba}, viable parameter space of 
its mass and coupling still exists. 
In case of a scalar boson, we apply a similar  Lagrangian,
\begin{equation}
  \mathcal L_{S'}
\equiv
  g_{S'} Q_{\alpha \beta}
\left[
  \overline{\ell_{R \alpha}} \ell_{L \beta}
+ \overline{\nu^c_{L \alpha}} \nu_{L \beta}
\right] S' + h.c.
\end{equation}
Here the neutrino Yukawa coupling with the scalar $S'$ is of 
Majorana type, but the results do not depend on that. For comparison,
we keep the same charge assignment $Q_{\alpha \beta}$ as the $Z'$ case. 
The above two Lagrangians are implemented in CompHEP \cite{CompHEP}.

\begin{figure}[b]
\centering
\includegraphics[height=7cm,width=4.5cm,angle=-90]{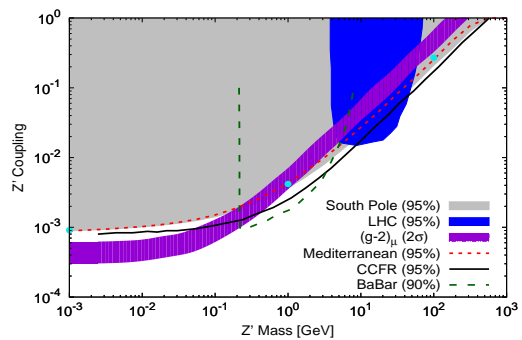}
\caption{The sensitivity to the $Z'$ mass and coupling at South Pole (PINGU+DeepCore+IceCube)
         Mediterranean (ORCA+ARCA), LHC \cite{Altmannshofer:2014pba}, 
         $(g-2)_\mu$ \cite{Altmannshofer:2014pba},
         CCFR \cite{Altmannshofer:2014pba}, and BaBar \cite{BaBar} experiments.}
\label{fig:chi2-Z'}
\end{figure}
\begin{figure}[h]
\centering
\includegraphics[height=7cm,width=4.5cm,angle=-90]{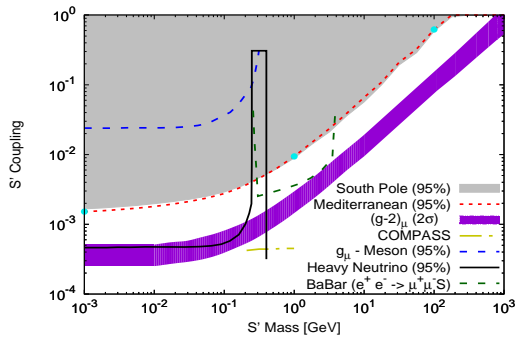}
\caption{The sensitivity to the $S'$ mass and coupling at South Pole (PINGU+DeepCore+IceCube),
         Mediterranean (ORCA+ARCA) and other experiments \cite{Batell:2016ove, Pasquini:2015fjv}.}
\label{fig:chi2-S'}
\end{figure}

We show the sensitivity to $M_{Z'}$ and $g_{Z'}$ in
\gfig{fig:chi2-Z'}. Below $M_{Z'} \sim 0.1\,\mbox{GeV}$, the curve 
is roughly flat while it grows fast for $M_{Z'} \gg 0.1\,\mbox{GeV}$. This is because
the $Z'$ contribution contains a propagator $1/(q^2_{Z'} - M^2_{Z'})$, where
$q_{Z'}$ is the $t$-channel momentum transfer of $Z'$. For light $Z'$, $q^2_{Z'} \gg M^2_{Z'}$,
the propagator is approximately 
$1/q^2_{Z'}$, where $q_{Z'} \gtrsim 1 \, \mbox{GeV}$ is the typical momentum transfer,
and hence insensitive to 
$M^2_{Z'}$. On the other hand, for heavy $Z'$,
$q^2_{Z'} \ll M^2_{Z'}$, the propagator is roughly $1/M^2_{Z'}$ and hence suppresses the rate 
with increasing $M_{Z'}$. 
The same reasoning applies for $S'$ as shown in \gfig{fig:chi2-S'}. In both cases, the final sensitivities 
for the combinations PINGU+DeepCore and ORCA+ARCA are very similar in the plots. 

For comparison, in \gfig{fig:chi2-Z'} and \gfig{fig:chi2-S'} we also show  constraints
from other experiments. Although some of them are more stringent, such as $(g-2)_\mu$
measurement, they cannot directly
confront and exclude the neutrino trident production. The interpretation of experimental
data relies on theoretical assumption. The $(g-2)_\mu$ bound is based on the assumption
that there is only $Z'$ or $S'$ contribution. Nevertheless, if there is something more
that can also contribute to $(g-2)_\mu$ but not neutrino trident production, the $(g-2)_\mu$
constraint on $g_{Z'/S'}$ and $M_{Z'/S'}$ can be easily evaded. In addition, although the
CCFR trident measurement $\sigma_{\rm CCFR}/\sigma_{\rm SM} = 0.82 \pm 0.28$, in contrast
to the $11\%$ ($16\%$) at the South Pole (Mediterranean) combination, is claimed to give
better sensitivity on $Z'$, it might come from the fact that the central value deviates
from the SM prediction.

\begin{figure}[h]
\centering
\includegraphics[height=4.1cm,width=4.43cm,angle=-90]{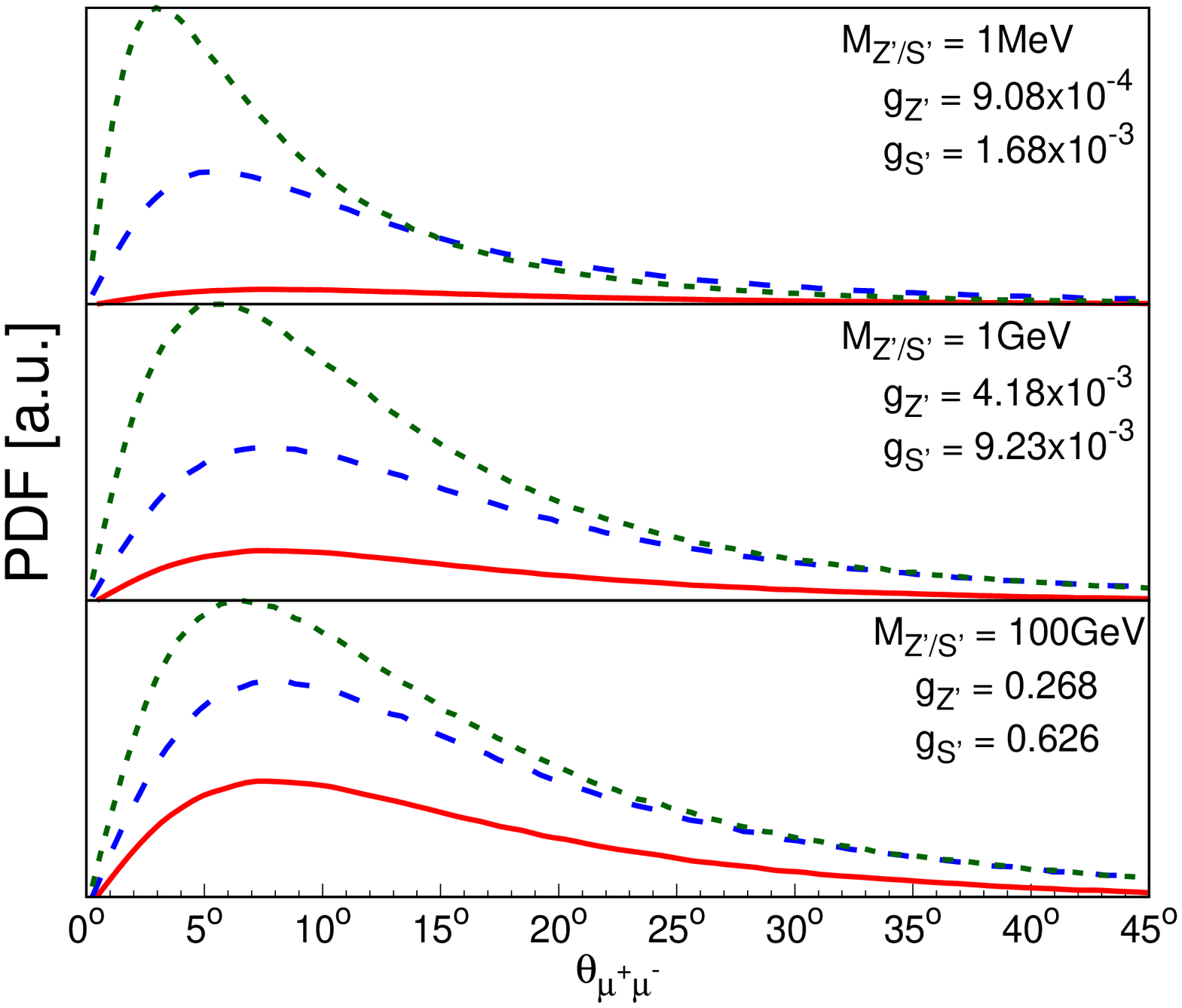}
\includegraphics[height=4.1cm,width=4.43cm,angle=-90]{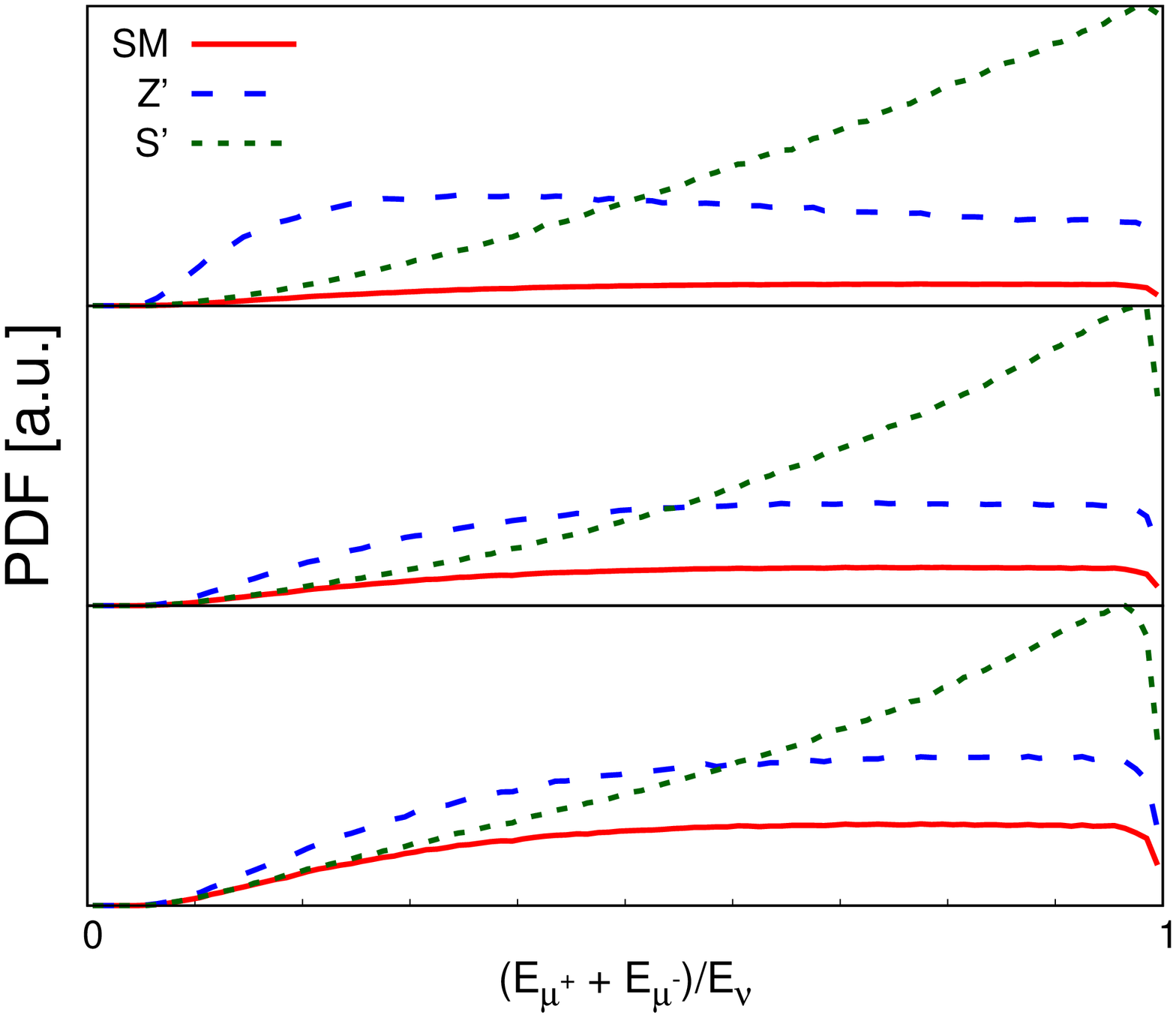}
\vspace{2mm}
\caption{The distributions of the opening angle $\theta_{\mu^+ \mu^-}$ (left) and total energy (right) of
         the muon pair produced in the neutrino trident production at oxygen nuclei.}
\label{fig:cuts}
\end{figure}

In the current analysis we only use the total event number.
There are several ways to further enhance the sensitivity. 
First, using differential distributions can significantly enhance
the sensitivity of identifying NP from the SM, especially when the total
cross section is comparable but differential distributions are
different. We show the distributions of the opening angle $\theta_{\mu^+ \mu^-}$ and
energy sum $E_{\mu^+} + E_{\mu^-}$ the muon
pair in \gfig{fig:cuts} for illustration. 
Secondly, not just fully contained events can contribute. Those events being produced
outside and going through or stop inside the detector can also be identified with two
coincident tracks. Although not fully contained, the muon energy can be reconstructed
according to the radiation pattern along the way, which is the so-called
Edepillim \cite{Edepillim} algorithm. The Edepillim algorithm can also identify
overlapping double muon tracks.
Finally, final-state leptons of not only muon flavor but also other combinations can be
used to extract information on NP. With all improvements added, we can expect much
better sensitivities from atmospheric neutrino experiments than the ones shown in
\gfig{fig:chi2-Z'} and \gfig{fig:chi2-S'}.

{\it \bf Conclusion} --
We propose to use the free source of atmospheric neutrinos to probe neutrino physics,
here in the form of vector and scalar bosons with coupling to muon neutrinos. 
The huge flux of atmospheric neutrinos and effective volume of large Cherenkov detectors
are of great advantage to
observe first the rare SM process of neutrino trident production, and also look
for possible deviations arising from new physics. A pair of final-state muons can leave distinctive
double tracks in the ice/water Cherenkov detector. Our analysis demonstrates that, in addition 
to pursuing their standard physics program, 
such experiments can make very useful contributions in constraining new physics without
changing their configuration. This essentially turns neutrino telescope into neutrino
collider. The original scripts for running CompHEP in batch mode can be downloaded
from the \href{https://gitlab.com/gesf/NuTrident\_CompHEP}{NuTrident\_CompHEP} project
at GitLab. \\

{\it Acknowledgements} --
SFG would like to thank 
Wolfgang Altmannshofer,
Vedran Brdar,
J\"{u}rgen Brunner,
Romulus Godang,
Francis Halzen,
Jannik Hofest\"{a}dt,
Morihiro Honda,
Clancy James,
David McKeen,
Antoine David Kouchner,
Pedro Pasquini,
Sally Robertson,
Carsten Rott
for useful discussions and kind help.
WR thanks Julian Heeck for helpful comments. 
This work was supported by the DFG with grant RO 2516/6-1 in the Heisenberg Programme (WR).

\end{document}